\documentclass[10pt]{article}
\pdfoutput=1
\usepackage{geometry}
\usepackage{amsfonts}
\usepackage{amsmath}
\usepackage{amssymb}
\usepackage{mathptmx}
\usepackage{graphicx}
\usepackage{hyperref}
\usepackage{cancel}
\usepackage{textcomp}
\usepackage{bm}
\usepackage{times}
\usepackage{epsfig}
\usepackage{xcolor}
\usepackage{colortbl}
\usepackage{slashed}
\usepackage{booktabs}
\usepackage{multirow}
\usepackage{latexsym}
\usepackage{verbatim}
\usepackage{longtable}
\usepackage{wasysym}
\definecolor{mygray}{gray}{.95}
\usepackage[title]{appendix}
\allowdisplaybreaks[4]
\usepackage{capt-of}

\newcommand{\Tr}{\rm Tr}

\newcommand{\calL}{{\cal L}}
\newcommand{\calM}{{\cal M}}
\newcommand{\calO}{{\cal O}}

\newcommand{\calX}{{\cal X}}
\newcommand{\calY}{{\cal Y}}

\newcommand{\TeV}{\rm TeV}
\newcommand{\GeV}{\rm GeV}

\newcommand{\eV}{\rm eV}

\newcommand{\chpt}{\chi{\rm PT}}

\begin{document}
\baselineskip=16pt

\pagenumbering{arabic}

\vspace{1.0cm}

\begin{center}
{\Large\sf Effective field theory approach to lepton number violating $\tau$ decays}
\\[10pt]
\vspace{.5 cm}

{Yi Liao~$^{a,c}$\footnote{liaoy@nankai.edu.cn},~
	Xiao-Dong Ma~$^{b}$\footnote{maxid@phys.ntu.edu.tw},~
	Hao-Lin Wang~$^{a}$\footnote{whaolin@mail.nankai.edu.cn} }

{$^a$~School of Physics, Nankai University, Tianjin 300071, China
	\\
	$^b$ Department of Physics, National Taiwan University, Taipei 10617, Taiwan
	\\
	$^c$ Center for High Energy Physics, Peking University, Beijing 100871, China}

\vspace{2.0ex}

{\bf Abstract}
\end{center}

We continue our endeavor to investigate lepton number violating (LNV) processes at low energies in the framework of effective field theory (EFT).  In this work we study the LNV tau decays $\tau^+\rightarrow \ell^-P_i^{+}P_j^{+}$, where $\ell=e,~\mu$ and $P^+_{i,j}$ denote the lowest-lying charged pseudoscalars $\pi^+,~K^+$. We analyze the dominant contributions in a series of EFTs from high to low energy scales, namely the standard model EFT (SMEFT), the low-energy EFT (LEFT), and the chiral perturbation theory ($\chpt$). The decay branching ratios are expressed in terms of the Wilson coefficients of dimension-five and -seven operators in SMEFT and the hadronic low-energy constants. These Wilson coefficients involve the first and second generations of quarks and all generations of leptons; thus, they cannot be explored in low-energy processes such as nuclear neutrinoless double beta decay or LNV kaon decays. Unfortunately, the current experimental upper bounds on the branching ratios are too weak to set useful constraints on these coefficients. Alternatively, if we assume the new physics scale is larger than 1~TeV, the branching ratios are well below the current experimental bounds. We also estimate the hadronic uncertainties incurred in applying $\chpt$ to $\tau$ decays by computing one-loop chiral logarithms and attempt to improve the convergence of chiral perturbation by employing dispersion relations in the short-distance part of the decay amplitudes.

\newpage

%%%%%%%%%%%%%%%
\section{Introduction}
\label{intro}
%%%%%%%%%%%%%%%
Whilst neutrino oscillation experiments provide definite evidence for the existence of neutrino mass, its origin and the nature of neutrinos remain mysterious. As neutral fermions, neutrinos might well be Majorana particles as it naturally happens in conventional seesaw mechanisms of neutrino mass generation thus resulting in lepton number violation. In the meantime, one searches for new heavy particles presumably involved in Majorana mass generation at high energy colliders through like-sign dilepton production; hence, it is important to explore lepton number violating (LNV) signals in precision low energy processes. The nuclear neutrinoless double beta decay ($0\nu\beta\beta$) has so far provided the largest data sample and set the strongest constraint on lepton number violation in the first generation of leptons and quarks~\cite{KamLAND-Zen:2016pfg,Gomez-Cadenas:2019sfa,Agostini:2018tnm}. Under these circumstances, we should keep conscious that new physics might first reveal itself in processes involving heavier leptons and quarks as the usual wisdom indicates. Indeed, in recent years, the LNV decays of mesons such as $K^\pm,~D^\pm,~D^\pm_s,~B^\pm$ and the $\tau$ lepton have been continuously searched for in many experiments, including LHCb~\cite{Aaij:2012zr, Aaij:2011ex,Aaij:2013sua,Aaij:2014aba}, BaBar~\cite{Lees:2011hb, BABAR:2012aa, Lees:2013gdj}, Belle~\cite{Seon:2011ni,Miyazaki:2012mx}, CLEO~\cite{Rubin:2010cq} and others~\cite{Amhis:2016xyh,CortinaGil:2019dnd,Appel:2000tc,Kodama:1995ia}, and significantly improved constraints on some of the decays are expected in upgraded or proposed experiments~\cite{Chun:2019nwi,Perez:2019cdy}. From the theoretical point of view, it is advantageous that we avoid complicated nuclear physics in these decays, although we have to cope with hadronic uncertainties in most cases.

In previous publications~\cite{Liao:2019gex,Liao:2020roy}, we investigated the LNV decays $K^\pm\rightarrow\pi^\mp \ell^\pm_\alpha \ell^\pm_\beta$ (with $\ell_{\alpha,\beta}^\pm=e^\pm,~\mu^\pm$) completely in the framework of effective field theory (EFT), including both short-distance (SD) and long-distance (LD) contributions. Note that $K^\pm$ are the lightest hadrons whose decays could violate lepton number conservation in the charged lepton sector. In this work, we expand our study to the single charged lepton which can decay hadronically while violating lepton number conservation, i.e., the three-body $\tau$ lepton decays, $\tau^\pm\rightarrow \ell^\mp_\alpha P_i^\pm P_j^\pm$, with $P_{i,j}^\pm=\pi^\pm,~K^\pm$. The best upper limits on the branching ratios of these decays were obtained from the Belle experiment~\cite{Miyazaki:2012mx} as
\begin{align}
	\label{ul1}
	&\mathcal{B}(\tau^-\rightarrow e^+\pi^-\pi^-)<2.0\times10^{-8},
	&&\mathcal{B}(\tau^-\rightarrow \mu^+ \pi^-\pi^-)<3.9\times10^{-8},\\
	\label{ul2}
	&\mathcal{B}(\tau^-\rightarrow e^+K^-K^-)<3.3\times10^{-8},
	&&\mathcal{B}(\tau^-\rightarrow \mu^+ K^-K^-)<4.7\times10^{-8},\\
	\label{ul3}
	&\mathcal{B}(\tau^-\rightarrow e^+K^-\pi^-)<3.2\times10^{-8},
	&&\mathcal{B}(\tau^-\rightarrow \mu^+ K^-\pi^-)<4.8\times10^{-8}.
\end{align}
These are expected to be improved in the Belle II experiment~\cite{Perez:2019cdy}. Though the bounds are approximately two orders of magnitude weaker than those on the LNV $K^\pm$ decays, they still provide unique information on lepton number violation involving the $\tau$ lepton and are therefore worth further exploration. We continue to work in the EFT framework. The most salient feature of the EFT approach is its universality. To study physics below the electroweak scale, we only have to assume whether there are any new and relatively light particles; meanwhile, different high-energy-scale physics is reflected in the Wilson coefficients in EFT at low energy.

This paper is organized as follows. By assuming no new particles to be lighter than the electroweak scale $\Lambda_{\rm EW}$, we start in Section~\ref{SMEFT} with the standard model EFT (SMEFT) whose dimension-5 (dim-5) and -7 operators provide the dominant effective LNV interactions. At the scale $\Lambda_{\rm EW}$ we perform matching calculations between the SMEFT and low energy EFT (LEFT) up to dim-9 operators in the latter; these are relevant to the decays under consideration. Then, in Section~\ref{chpt}, we study the chiral realization below the chiral symmetry breaking scale $\Lambda_\chi$ of the effective interactions in the LEFT, and we calculate the decay amplitudes. Next, in Section~\ref{disp}, we estimate the hadronic uncertainties produced by the relatively large mass of the $\tau$ lepton by computing one-loop chiral logarithms, and we attempt to improve the convergence of  the chiral perturbation using dispersion relations. Our master formulas for the decay branching ratios are presented in Section~\ref{constr}, together with the numerical estimates. We summarize our main results in Section~\ref{conclude}.

%%%%%%%%%%%%%%%
\section{SMEFT, LEFT, and their matching}
\label{SMEFT}
%%%%%%%%%%%%%%%

In light of the null result in searching for new particles of masses up to the TeV scale, it is plausible to assume that new physics appears at a scale $\Lambda_{\rm NP}$ well above the electroweak scale $\Lambda_{\rm EW}$ and that there are no new particles with a mass of order $\Lambda_{\rm EW}$ or below. Thus, we can establish an EFT, the SMEFT, between the two scales that is composed of the standard model (SM) fields and respects the SM gauge symmetries: $SU(3)_C\times SU(2)_L\times U(1)_Y$. Its Lagrangian is the SM Lagrangian $\calL_\textrm{SM}$ augmented by an infinite sum of effective interactions involving higher and higher dimensional operators, which are suppressed by increasingly large powers of $\Lambda_{\rm NP}$:
\begin{align}
	\calL_\textrm{SMEFT}=\calL_\textrm{SM}
	+\calL_5+\calL_6+\calL_7+\cdots.
\end{align}
Here, $\calL_5=C_{LH5}^{\alpha\beta}\calO_5^{\alpha\beta}$ contains the unique dim-5 Weinberg operator~\cite{Weinberg:1979sa},
\begin{align}
	\calO_5^{\alpha\beta}
	=\epsilon_{ij}\epsilon_{mn}(\overline{L^{C,i}_\alpha}L^m_\beta)
	H^jH^n,
\end{align}
which induces Majorana neutrino mass when the Higgs doublet field $H$ develops a vacuum expectation value. Here, $L_\beta$ refers to the left-handed doublet lepton field of flavor $\beta$, and $ijmn$ are the $SU(2)_L$ indices in the fundamental representation. $\calL_6$ collects the effective interactions of dim-6 operators~\cite{Buchmuller:1985jz,Grzadkowski:2010es}, and $\calL_7$ is a sum of dim-7 operators~\cite{Lehman:2014jma,Liao:2016hru}. For the decays under consideration here, LNV $\calL_5$ and $\calL_7$ have a dominant contribution. The dim-7 operators were first systematically studied in~\cite{Lehman:2014jma}; their basis was established in~\cite{Liao:2016hru} by removing redundancies and further refined in~\cite{Liao:2020roy} by making the flavor symmetries manifest. For an earlier survey on LNV effective operators, see~\cite{deGouvea:2007qla}. The dim-7 operators that violate lepton number conservation by two units but conserve baryon number~\cite{Liao:2020roy} are listed in Table~\ref{tabSMEFT}.

\begin{table}
	\centering
	\begin{tabular}{|c| l |c| l |}
		\hline
		$\psi^2H^4$
		& $\mathcal{O}_{LH}=\epsilon_{ij}\epsilon_{mn}(\overline{L^{C,i}}L^m)H^jH^n(H^\dagger H)$
		& \multirow{5}{*}{$\psi^4H$ }
		& $\mathcal{O}_{\overline{e}LLLH}=\epsilon_{ij}\epsilon_{mn}(\overline{e}L^i)(\overline{L^{C,j}}L^m)H^n$
		\\\cline{1-2}
		$\psi^2H^3D$
		& $\mathcal{O}_{LeHD}=\epsilon_{ij}\epsilon_{mn}(\overline{L^{C,i}}\gamma_\mu e)H^j(H^miD^\mu H^n)$
		&
		 &$\mathcal{O}_{\overline{d}QLLH1}=\epsilon_{ij}\epsilon_{mn}(\overline{d}Q^i)(\overline{L^{C,j}}L^m)H^n$
		\\\cline{1-2}
		\multirow{2}{*}{$\psi^2H^2X$ }
		 &$\mathcal{O}_{LHB}=g_1\epsilon_{ij}\epsilon_{mn}(\overline{L^{C,i}}\sigma_{\mu\nu}L^m)H^jH^nB^{\mu\nu}$
		&
		& $\mathcal{O}_{\overline{d}QLLH2}=\epsilon_{ij}\epsilon_{mn}(\overline{d}\sigma_{\mu\nu}Q^i)(\overline{L^{C,j}}\sigma^{\mu\nu} L^m)H^n$
		\\
		&$\mathcal{O}_{LHW}=g_2\epsilon_{ij}(\epsilon \tau^I)_{mn}(\overline{L^{C,i}}\sigma_{\mu\nu}L^m)H^jH^nW^{I\mu\nu}$
		&
		& $\mathcal{O}_{\overline{d}uLeH}=\epsilon_{ij}(\overline{d}\gamma_\mu u)(\overline{L^{C,i}}\gamma^\mu e)H^j$
		\\\cline{1-2}
		\multirow{2}{*}{$\psi^2H^2D^2$ }
		& $\mathcal{O}_{LDH1}=\epsilon_{ij}\epsilon_{mn}(\overline{L^{C,i}}\overleftrightarrow{D}_\mu L^j)(H^mD^\mu H^n)$
		&
		& $\mathcal{O}_{\overline{Q}uLLH}=\epsilon_{ij}(\overline{Q}u)(\overline{L^{C}}L^i)H^j$
		\\\cline{3-4}
		& $\mathcal{O}_{LDH2}=\epsilon_{im}\epsilon_{jn}(\overline{L^{C,i}}L^j)(D_\mu H^m D^\mu  H^n)$
		& $\psi^4D$
		& $\mathcal{O}_{\overline{d}uLDL}=\epsilon_{ij}(\overline{d}\gamma_\mu u)(\overline{L^{C,i}}i\overleftrightarrow{D}^\mu L^j)$
		\\\hline
	\end{tabular}
	\caption{Basis of dim-7 LNV but baryon number conserving operators in SMEFT. $L,~Q$ are the left-handed lepton and quark doublet fields, respectively; $u,~d,~e$ are the right-handed up-type quark, down-type quark and charged lepton singlet fields, respectively; and $H$ denotes the Higgs doublet. $D_\mu$ is defined for the gauge symmetries $SU(3)_C\times SU(2)_L\times U(1)_Y$, and $D^\mu H^n$ is understood as $(D^\mu H)^n$.}
	\label{tabSMEFT}
\end{table}

The SM electroweak symmetries are spontaneously broken into $U(1)_{\rm EM}$ by the vacuum expectation value of the Higgs field $\langle H\rangle=(0,1)^Tv/\sqrt{2}$, which defines the electroweak scale $\Lambda_\textrm{EW}$. Upon integrating out the heavy particles in SMEFT (i.e., the Higgs, $W^\pm$, and $Z$ bosons and the top quark), we arrive at the LEFT for the remaining SM particles; this is another infinite sum of effective interactions involving higher and higher dimensional operators suppressed by increasingly large powers of $\Lambda_\textrm{EW}$. For dim-6 and dim-7 operators in the LEFT, we adopt the basis given in~\cite{Jenkins:2017jig} and~\cite{Liao:2020zyx}, respectively. The short-distance contribution to the $\tau$ decays under consideration arises from dim-9 LNV operators involving four quark and two lepton fields, whose basis was determined in~\cite{Liao:2019gex}. All the operators relevant to our discussion here are collected in Table~\ref{tabSM2LEFT}. They are classified according to the types of contributions finally entering the $\tau$ decays: Majorana neutrino mass insertion (MM), long-distance (LD), and short-distance (SD), see Fig.~\ref{fig1}. Here long distance refers to the exchange of a light neutrino and short distance indicates contact interactions between the initial and final particles in the $\tau$ decays.

The two EFTs, the SMEFT at scales above $\Lambda_\textrm{EW}$ and the LEFT below, are related by matching conditions at scale $\Lambda_\textrm{EW}$. Here, we perform tree-level matching; the results are shown in Table~\ref{tabSM2LEFT}. Our convention in the LEFT is that we work with mass eigenstate fields of quarks and charged leptons but with flavor eigenstate fields of neutrinos because the neutrino mass appears only in the form of a matrix in flavor space. Some operators (e.g., dim-7 tensor operators) that generically exist in the LEFT are not induced at this level from the SMEFT. Some other operators that are not induced at the matching scale $\Lambda_\textrm{EW}$ are however generated at lower scales from other operators by renormalization-group running effects (e.g., the dim-9 operator $\calO_{prst}^{LRRL,S/P}$). Some of dim-7 SMEFT operators ($\calO_{\bar{Q}uLLH}$, $\calO_{\bar{d}QLLH1}$, $\calO_{\bar{d}QLLH2}$, $\calO_{LeHD}$, and $\calO_{\bar{d}uLeH}$) induce LNV dim-6 LEFT operators that involve a charged lepton, a neutrino, and a quark bilinear. These operators supposedly generate the leading contributions in Fig.~\ref{fig1}~(b). The other set of operators in SMEFT ($\calO_{LDH1}$, $\calO_{LHW}$, and $\calO_{\overline{d}uLDL}$) generates dim-7 LEFT operators that carry an additional covariant derivative $D_\mu$, which would contribute at the next-to-leading order in Fig.~\ref{fig1}~(b). Finally, among many possible LNV dim-9 LEFT operators involving four quarks and two charged leptons~\cite{Liao:2019gex}, only a few can be induced from dim-7 operators in SMEFT ( i.e., $\calO_{LHW}$, $\calO_{\overline{d}uLDL}$, $\calO_{LDH1}$, and $\calO_{LDH2}$). This significantly simplifies calculations.

\begin{table}%[!ht]
	\centering
	\begin{tabular}{|c| l | l |}
		\hline
		Types and dim.
		& Operators in LEFT
		& Matching LEFT (left) with SMEFT (right) at $\Lambda_\textrm{EW}$
		\\
		\hline
		MM: dim-3 & ${\cal L}_{\rm M}=-\frac{1}{2}m_{\alpha\beta}\overline{\nu^C_\alpha}\nu_\beta$ & $m_{\alpha\beta}=-v^2C_{LH5}^{\alpha\beta*}-\frac{1}{2} v^4C_{LH}^{\alpha\beta*}$
		\\
		\hline
		\multirow{5}{*}{LD: dim-6}
		& $\calO^{RL,S}_{pr\alpha\beta}
		=(\overline{u_R^p}d_L^r)(\overline{\ell_{L\alpha}}\nu^C_\beta)$
		& $C^{RL,S}_{pr\alpha\beta}
		=\frac{v}{\sqrt{2}}V_{wr}C_{\bar{Q}uLLH}^{wp\alpha\beta*}$
		\\
		& $\calO^{LR,S}_{pr\alpha\beta}
		=(\overline{u_L^p}d_R^r)(\overline{\ell_{L\alpha}}\nu^C_\beta)$
		& $C^{LR,S}_{pr\alpha\beta}
		=\frac{v}{\sqrt{2}}C_{\bar{d}QLLH1}^{rp\alpha\beta*}$
		\\
		& $\calO^{LL,V}_{pr\alpha\beta}
		=(\overline{u_L^p}\gamma_\mu d_L^r)
		(\overline{\ell_{R\alpha}}\gamma^\mu \nu^C_\beta)$
		& $C^{LL,V}_{pr\alpha\beta}
		=\frac{v}{\sqrt{2}}V_{pr}C_{LeHD}^{\beta\alpha *}$
		\\
		& $\calO^{RR,V}_{pr\alpha\beta}
		=(\overline{u_R^p}\gamma_\mu d_R^r)
		(\overline{\ell_{R\alpha}}\gamma^\mu \nu^C_\beta)$
		& $C^{RR,V}_{pr\alpha\beta}
		=\frac{v}{\sqrt{2}}C_{\bar{d}uLeH}^{r p\beta\alpha*}$
		\\
		& $\calO^{LR,T}_{pr\alpha\beta}
		=(\overline{u_L^p}\sigma_{\mu\nu}d_R^r)
		(\overline{\ell_{L\alpha}}\sigma^{\mu\nu}\nu^C_\beta)$
		& $C^{LR,T}_{pr\alpha\beta}
		=\frac{v}{\sqrt{2}}C_{\bar{d}QLLH2}^{rp\alpha\beta*}$
		\\
		\hline
		\multirow{2}{*}{LD: dim-7}
		& $\calO^{LL,VD}_{pr\alpha\beta}
		=(\overline{u_L^p}\gamma_\mu d_L^r)
		(\overline{\ell_{L\alpha}}i\overleftrightarrow{D}^\mu\nu^C_\beta)$
		& $C^{LL,VD}_{pr\alpha\beta}
		=-V_{pr}\left( 4C_{LHW}^{\beta\alpha*}
		+2C_{LDH1}^{\alpha\beta*}\right)$
		\\
		& $\calO^{RR,VD}_{pr\alpha\beta}
		=(\overline{u_R^p}\gamma_\mu d_R^r)
		(\overline{\ell_{L\alpha}}i\overleftrightarrow{D}^\mu\nu^C_\beta)$
		& $C^{RR,VD}_{pr\alpha\beta}
		=2C_{\bar{d}uLDL}^{rp\alpha\beta*}$
		\\
		\hline
		\multirow{4}{*}{SD: dim-9}
		& $\calO_{prst,\alpha\beta}^{LLLL,S/P}
		=(\overline{u_L^p}\gamma^\mu d_L^r)
		[\overline{u_L^s}\gamma_\mu d_L^t]j^{\alpha\beta}_{(5)}$		& $C_{prst,\alpha\beta}^{LLLL,S/P}
		=-2\sqrt{2}G_FV_{pr}V_{st}$
		\\
		&
		& $\hspace{1.5cm}\times\left( C_{LHW}^{\alpha\beta*}+C_{LHW}^{\beta\alpha*}
		+C_{LDH1}^{\alpha\beta*}
		+{1\over 2}C_{LDH2}^{\alpha\beta*}\right)$
		\\
		& $\calO_{prst,\alpha\beta}^{LRRL, S/P}
		=(\overline{u_L^p}d_R^r)
		[\overline{u_R^s}d_L^t]j^{\alpha\beta}_{(5)}$
		& $C_{prst,\alpha\beta}^{LRRL,S/P}=0$
		\\
		& $\tilde{\calO}_{prst,\alpha\beta}^{LRRL,S/P}
		=(\overline{u_L^p}d_R^r]
		[\overline{u_R^s}d_L^t)j^{\alpha\beta}_{(5)}$
		& $\tilde{C}_{prst,\alpha\beta}^{LRRL~S/P}
		=-4\sqrt{2}G_FV_{pt}C_{\bar{d}uLDL}^{rs\alpha\beta*}$
		\\
		\hline
	\end{tabular}
	\caption{Relevant LEFT operators (middle column) and their Wilson coefficients (right column) obtained from matching at the scale $\Lambda_\textrm{EW}$ to those of SMEFT dim-5 and dim-7 effective interactions. The Wilson coefficients carry identical indices as corresponding operators in both SMEFT and LEFT. Here, $j^{\alpha\beta}=(\overline{\ell_{\alpha}} \ell_{\beta}^C)$, $j^{\alpha\beta}_5=(\overline{\ell_{\alpha}}\gamma_5 \ell_{\beta}^C)$, $D_\mu$ refers to gauge symmetries $SU(3)_C\times U(1)_\textrm{EM}$, and $(\cdots)$ and $[\cdots]$ denote two color contractions.}
	\label{tabSM2LEFT}
\end{table}

We aim to calculate the $\tau$ decays at even lower energies; hence, we match the effective operators in LEFT to those in $\chpt$ at the scale $\Lambda_\chi=4\pi F_\pi \approx 1.2~\rm GeV$. For this, we first must perform the renormalization-group (RG) running of the Wilson coefficients in LEFT, using the matching results in Table \ref{tabSM2LEFT} as the initial conditions. At this stage of study, it suffices to include the one-loop QCD effects previously computed in~\cite{Liao:2019gex,Liao:2020roy}. For dim-6 scalar and tensor operators, the RG results are
\begin{align}
	&C^S(\Lambda_\chi)=1.656 C^S(\Lambda_\text{EW}),~C^S\in \{C_{pr\alpha\beta}^{RL,S},C_{pr\alpha\beta}^{LR,S}\},
	\nonumber
	\\
	&C_{pr\alpha\beta}^{LR,T}(\Lambda_\chi)
	=0.845C_{pr\alpha\beta}^{LR,T}(\Lambda_\text{EW}).
	\label{rge11}
\end{align}
The other dim-6 and -7 operators involve a quark vector current and therefore do not run owing to the QCD Ward identity. As mentioned above, the renormalization of dim-9 operators induces operator mixing as well as running:
\begin{align}
	C_{uiuj}^{LLLL,S/P}(\Lambda_\chi)=&
	0.78 C_{uiuj}^{LLLL,S/P}(\Lambda_\text{EW}),
	\label{eqr1}
	\\
	\tilde{C}_{uiuj}^{LRRL,S/P}(\Lambda_\chi)=&
	0.88\tilde{C}_{uiuj}^{LRRL,S/P}(\Lambda_\text{EW}),
	\label{eqr2}
	\\
	C_{uiuj}^{LRRL,S/P}(\Lambda_\chi)=&
	0.62\tilde{C}_{uiuj}^{LRRL,S/P}(\Lambda_\text{EW}).
	\label{eqr3}
\end{align}
Here, the RG running from higher to lower scales represents a mild suppression, with the exception of the dim-6 scalar operators which are enhanced.

%%%%%%%%%%%%%%%
\section{Decay amplitudes in $\chpt$}
\label{chpt}
%%%%%%%%%%%%%%%

At the chiral symmetry breaking scale $\Lambda_{\chi}$, the approximate chiral symmetry $G=SU(3)_L\times SU(3)_R$ for the $q=u,~d,~s$ quarks in the QCD Lagrangian is spontaneously broken to $H=SU(3)_V$ by the quark condensate $\langle 0|\bar qq|0\rangle=-3BF_0^2$. This generates an octet of Nambu-Goldstone (NG) bosons living in the coset space $G/H$. When the small quark masses are taken into account, they become the so-called pseudo-NG bosons and can be identified with the lowest-lying octet of pseudoscalar mesons $\pi^\pm,~\pi^0,~K^\pm,~K^0,~\overline{K^0},~\eta$. Their strong interactions at low energies are best described by chiral perturbation theory ($\chpt$)~\cite{Gasser:1983yg,Gasser:1984gg}. The framework of $\chpt$ is flexible enough to describe additional interactions of the mesons inherited from effective interactions of light quarks in the LEFT. This is exactly what we want to do next for the nonperturbative matching at the scale $\Lambda_\chi$, where the light quark degrees of freedom give way to the mesons. This is based upon the analysis of linear versus nonlinear realizations of the chiral symmetry. For effective interactions involving a single quark bilinear (i.e., the dim-6 and -7 operators in Table~\ref{tabSM2LEFT}), the products of other fields multiplying the bilinear are treated as external sources coupled to light quarks in the QCD Lagrangian. For effective interactions with more quark bilinears, we can apply the technique of spurion analysis, which has been elaborated upon in~\cite{Graesser:2016bpz,Liao:2019gex} in the context of the dim-9 operators in Table~\ref{tabSM2LEFT}.

\begin{figure}%[!h]
	\centering.
	\includegraphics[width=15cm]{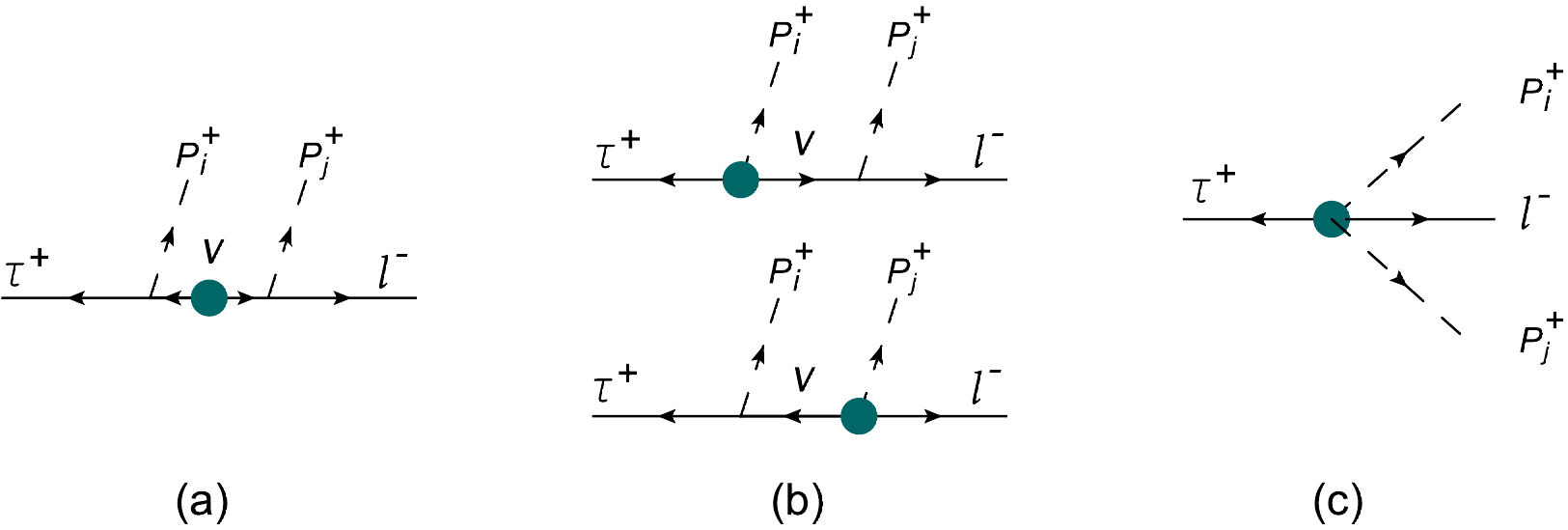}
	\caption{Feynman diagrams for decay $\tau^+\rightarrow \ell^-P_i^{+}P_j^{+}$ in $\chpt$. The heavy blob denotes effective LNV interactions, and the arrow on the lepton (meson) line indicates lepton number (positive charge) flow. Crossing diagrams in (a,~b) are not shown.}
	\label{fig1}
\end{figure}

In $\chpt$, the meson fields are parameterized by~\cite{Coleman:1969sm,Callan:1969sn}
\begin{align}
	\Sigma(x)&=\exp\left(\frac{i\sqrt{2}\Pi(x)}{F_0}\right), &
	\Pi&=\begin{pmatrix}
		\frac{\pi^0}{\sqrt{2}}+\frac{\eta}{\sqrt{6}} & \pi^+ & K^+
		\\
		\pi^- & -\frac{\pi^0}{\sqrt{2}}+\frac{\eta}{\sqrt{6}} & K^0
		\\
		K^- & \bar{K}^0 & -\sqrt{\frac{2}{3}}\eta
	\end{pmatrix},
\end{align}
where $F_0$ is the decay constant in the chiral limit. The leading-order $\calO(p^2)$ Lagrangian incorporating the scalar and pseudoscalar ($\chi$) and vector ($l_\mu,~r_\mu$) external sources is given by
\begin{eqnarray}
	\label{p2l}
	\mathcal{L}^{(2)}_{\chpt}
	=\frac{F_0^2}{4}{\Tr}\left(D_\mu \Sigma (D^\mu \Sigma)^\dagger \right)+\frac{F_0^2}{4}{\Tr} \left(\chi \Sigma^\dagger +\Sigma\chi^\dagger \right),
\end{eqnarray}
where $D_\mu \Sigma=\partial_\mu \Sigma-i l_\mu \Sigma+i \Sigma r_\mu$. The external tensor sources ($t_l^{\mu\nu},~t_r^{\mu\nu}$) first appear at $\calO(p^4)$~\cite{Cata:2007ns}:
\begin{eqnarray}
	\label{p4l}
	\mathcal{L}^{(4)}_{\chpt}\supset i\Lambda_2 {\Tr}\left(t_l^{\mu\nu}(D_\mu \Sigma)^\dagger U (D_\nu U)^\dagger+t_r^{\mu\nu}D_\mu UU^\dagger D_\nu U\right).
\end{eqnarray}
By inspecting the dim-6 and dim-7 operators in Table~\ref{tabSM2LEFT} (which appear as additional terms in the QCD Lagrangian when multiplied by their Wilson coefficients), we can read off the external sources relevant to the decays under consideration, as
\begin{eqnarray}
	(l^\mu)_{ui}&=&
	-2\sqrt{2}G_FV_{ui}
	(\overline{\ell_{L\alpha}}\gamma^\mu\nu_\alpha) +C^{LL,V}_{ui\alpha\beta}
	(\overline{\ell_{R\alpha}}\gamma^\mu\nu^C_\beta)
	+C^{LL,VD}_{ui\alpha\beta}(\overline{\ell_{L\alpha}}
	i\overleftrightarrow{D}^\mu\nu^C_\beta)+\cdots,
	\\
	(r^\mu)_{ui}&=&
	C^{RR,V}_{ui\alpha\beta}
	(\overline{\ell_{R\alpha}}\gamma^\mu \nu^C_\beta)
	+C^{RR,VD}_{ui\alpha\beta}(\overline{\ell_{L\alpha}}
	i\overleftrightarrow{D}^\mu\nu^C_\beta)+\cdots,
	\\
	(\chi^\dagger)_{ui}&=& 2BC^{RL,S}_{ui\alpha\beta}(\overline{\ell_{L\alpha}}\nu^C_\beta)
	+\cdots,
	\\
	(\chi)_{ui}&=& 2BC^{LR,S}_{ui\alpha\beta}
	(\overline{\ell_{L\alpha}}\nu^C_\beta)+\cdots,
	\\
	(t_l^{\mu\nu})_{ui}&=& C^{LR,T}_{ui\alpha\beta}
	(\overline{\ell_{L\alpha}}\sigma^{\mu\nu}\nu^C_\beta)+\cdots,
	\\
	(t_r^{\mu\nu})_{ui}&=&0,
\end{eqnarray}
where $i=d,~s$ (or $i=2,~3$ when labeling the $\Sigma$ matrix indices), $V$ is the Cabibbo-Kobayashi-Maskawa (CKM) matrix, and the ellipses denote irrelevant terms. These source terms are the LNV interactions of the mesons with a charged lepton--neutrino pair, and they produce the diagrams in Fig.~\ref{fig1}~(b) when the neutrino field is contracted by a usual interaction vertex which defines the meson decay constant [see the first term in Eq.~\eqref{llong}]. Because each tensor term includes at least two mesons, it only contributes to the four-body decays of the $\tau$ lepton. We reserve this more complicated case for the future study, because the resonances must be explicitly included. Finally, when both neutrinos in the mass term $\calL_\textrm{M}$ are contracted to two usual vertices mentioned above, we obtain the diagram in Fig.~\ref{fig1}~(a).

Using all of the above details, we can now write down the interactions entering the LD contribution to the $\tau$ decays,
\begin{eqnarray}
	\calL^{(2)}_{\chpt}&\supset&
	F_0G_F\left(V_{ud}\partial_\mu \pi^- +V_{us}\partial_\mu K^-\right)\left(\overline{\ell_{L\alpha}}\gamma^\mu\nu_\alpha\right)
	\nonumber
	\\
	&&+F_0\Big[iB\left( c_{\pi 1}^{\alpha\beta} \pi^-+c_{K1}^{\alpha\beta}K^-\right)
	\left(\overline{\ell_{L\alpha}}\nu^C_\beta\right)
	-\left( c_{\pi 2}^{\alpha\beta}\partial_\mu \pi^-+c_{K2}^{\alpha\beta}\partial_\mu K^-\right)
	\left(\overline{\ell_{R\alpha}}\gamma^\mu\nu^C_\beta\right)
	\nonumber
	\\
	&&\hspace{.7cm}
	-\left(c_{\pi 3}^{\alpha\beta}\partial_\mu \pi^-+c_{K3}^{\alpha\beta}\partial_\mu K^-\right) \left(\overline{\ell_{L\alpha}}i\overleftrightarrow{D}^\mu \nu^C_\beta\right)\Big],
	\label{llong}
\end{eqnarray}
where the first term is the usual one and the others represent new LNV interactions. We have introduced the parameters
\begin{align}
	\label{cpi1}
	c_{P_i1}^{\alpha\beta}&=\frac{\sqrt{2}}{2} \left(C^{RL,S}_{ui\alpha\beta}-C^{LR,S}_{ui\alpha\beta}\right), &
	c_{P_i2}^{\alpha\beta}&=\frac{\sqrt{2}}{4}
	\left(C^{LL,V}_{ui\alpha\beta}-C^{RR,V}_{ui\alpha\beta} \right),
	&
	c_{P_i3}^{\alpha\beta}&=\frac{\sqrt{2}}{4}
	\left(C^{LL,VD}_{ui\alpha\beta}-C^{RR,VD}_{ui\alpha\beta} \right),
\end{align}
which are implicitly defined at the scale $\Lambda_\chi$, with $P_i=\pi,~K$ for $i=d,~s$. Employing the matching conditions in Table~\ref{tabSM2LEFT} at $\Lambda_\textrm{EW}$ and the one-loop QCD running effects in Eq.~\eqref{rge11} from $\Lambda_\textrm{EW}$ to $\Lambda_\chi$, we connect the above parameters at $\Lambda_\chi$ with the SMEFT Wilson coefficients defined at $\Lambda_{\rm EW}$:
\begin{align}
	\label{wcpi1}
	c_{P_i1}^{\alpha\beta}&=
	\frac{v}{2}(1.656)\calY^{\alpha\beta}_{P_i1},
	&
	c_{P_i2}^{\alpha\beta}&=
	\frac{v}{4}\calY^{\alpha\beta}_{P_i2},
	& c_{P_i3}^{\alpha\beta}&=
	\frac{\sqrt{2}}{4}\calY^{\alpha\beta}_{P_i3},
\end{align}
where
\begin{align}
	\calY^{\alpha\beta}_{P_i1}&=
	V_{wi}C_{\bar{Q}uLLH}^{w1\alpha\beta*}(\Lambda_{\rm EW})
	-C_{\bar{d}QLLH1}^{i1\alpha\beta*}(\Lambda_{\rm EW}),
	\nonumber
	\\
	\calY^{\alpha\beta}_{P_i2}&=
	V_{ui}C_{LeHD}^{\beta\alpha*}(\Lambda_{\rm EW})
	-C_{\bar{d}uLeH}^{i1\beta\alpha*}(\Lambda_{\rm EW}),
	\nonumber
	\\
	\calY^{\alpha\beta}_{P_i3}&=
	V_{ui}\left[4C_{LHW}^{\alpha\beta*}(\Lambda_{\rm EW})
	+C_{LDH1}^{\alpha\beta*}(\Lambda_{\rm EW})\right]
	-2C_{\bar{d}uLDL}^{i1\alpha\beta*}(\Lambda_{\rm EW}).
\label{Ydefined}
\end{align}

\begin{table}%[!ht]
	\centering
	\begin{tabular}{|c|c|c|c|}
		\hline
		Decays
		&LEFT operators
		& Chiral  irrep.
		& Hadronic operators
		\\ %%
		\hline
		&$\calO_{uiuj,\alpha\beta}^{LLLL,S/P}$
		& ${\bf 27}_L\times {\bf1}_R$
		& $\frac{5}{12}{g}_{27\times 1}F_0^4
		(\Sigma i\partial_\mu \Sigma^\dagger)_i^{~1}
		(\Sigma i\partial^\mu \Sigma^\dagger)_j^{~1}
		j^{\alpha\beta}_{(5)}$
		\\
		&${\calO}_{uiuj,\alpha\beta}^{LRRL,S/P}$
		& $ {\bf 8}_L\times {\bf 8}_R(a) $
		& $g_{8\times 8}^a\frac{F_0^4}{4}
		(\Sigma^\dagger)_i^{~1}(\Sigma)_j^{~1}
		j^{\alpha\beta}_{(5)}$
		\\
		$\tau^+\rightarrow \ell^-P^+_i P^+_j$
		&${\calO}_{ujui,\alpha\beta}^{LRRL,S/P}$
		& $ {\bf 8}_L\times {\bf 8}_R(a) $
		& $g_{8\times 8}^a\frac{F_0^4}{4}(\Sigma^\dagger)_j^{~1}
		(\Sigma)_i^{~1}j^{\alpha\beta}_{(5)}$
		\\
		&$\tilde{\calO}_{uiuj,\alpha\beta}^{LRRL,S/P}$
		& $ {\bf 8}_L\times {\bf 8}_R(b) $
		& $g_{8\times 8}^b\frac{F_0^4}{4}(\Sigma^\dagger)_i^{~1}
		(\Sigma)_j^{~1}j^{\alpha\beta}_{(5)}$
		\\
		&$\tilde{\calO}_{ujui,\alpha\beta}^{LRRL,S/P}$
		& $ {\bf 8}_L\times {\bf 8}_R(b) $
		& $g_{8\times 8}^b\frac{F_0^4}{4}(\Sigma^\dagger)_j^{~1}
		(\Sigma)_i^{~1}j^{\alpha\beta}_{(5)}$
		\\
		\hline %%
		\hline
		&$\calO_{udud}^{LLLL,S/P}$
		& ${\bf 27}_L\times {\bf1}_R$
		& $\frac{5}{12}{g}_{27\times 1}F_0^4
		(\Sigma i\partial_\mu \Sigma^\dagger)_2^{~1}
		(\Sigma i\partial^\mu \Sigma^\dagger)_2^{~1}(j/j_5)$
		\\
		$\tau^+\rightarrow \ell^-\pi^+\pi^+$
		&${\calO}_{udud}^{LRRL,S/P}$
		& $ {\bf 8}_L\times {\bf 8}_R(a) $
		& $g_{8\times 8}^a\frac{F_0^4}{4}(\Sigma^\dagger)_2^{~1}
		(\Sigma)_2^{~1}(j/j_5)$
		\\
		&$\tilde{\calO}_{udud}^{LRRL,S/P}$
		& $ {\bf 8}_L\times {\bf 8}_R(b) $
		& $g_{8\times 8}^b\frac{F_0^4}{4}(\Sigma^\dagger)_2^{~1}
		(\Sigma)_2^{~1}(j/j_5)$
		\\
		\hline
		\hline
		&$\calO_{usus}^{LLLL,S/P}$
		& ${\bf 27}_L\times {\bf1}_R$
		& $\frac{5}{12}{g}_{27\times 1}F_0^4
		(\Sigma i\partial_\mu \Sigma^\dagger)_3^{~1}
		(\Sigma i\partial^\mu \Sigma^\dagger)_3^{~1}(j/j_5)$
		\\
		$\tau^+\rightarrow \ell^-K^+K^+$
		&${\calO}_{usus}^{LRRL,S/P}$
		& $ {\bf 8}_L\times {\bf 8}_R(a) $
		& $g_{8\times 8}^a\frac{F_0^4}{4}(\Sigma^\dagger)_3^{~1}
		(\Sigma)_3^{~1}(j/j_5)$
		\\
		&$\tilde{\calO}_{usus}^{LRRL,S/P}$
		& $ {\bf 8}_L\times {\bf 8}_R(b) $
		& $g_{8\times 8}^b\frac{F_0^4}{4}(\Sigma^\dagger)_3^{~1}
		(\Sigma)_3^{~1}(j/j_5)$
		\\
		\hline
		\hline
		&$\calO_{udus}^{LLLL,S/P}$
		& ${\bf 27}_L\times {\bf1}_R$
		& $\frac{5}{12}{g}_{27\times 1}F_0^4
		(\Sigma i\partial_\mu \Sigma^\dagger)_2^{~1}
		(\Sigma i\partial^\mu \Sigma^\dagger)_3^{~1}(j/j_5)$
		\\
		&${\calO}_{udus}^{LRRL,S/P}$
		& $ {\bf 8}_L\times {\bf 8}_R(a) $
		& $g_{8\times 8}^a\frac{F_0^4}{4}(\Sigma^\dagger)_2^{~1}
		(\Sigma)_3^{~1}(j/j_5)$
		\\
		$\tau^+\rightarrow \ell^-K^+\pi^+$
		&${\calO}_{usud}^{LRRL,S/P}$
		& $ {\bf 8}_L\times {\bf 8}_R(a) $
		& $g_{8\times 8}^a\frac{F_0^4}{4}(\Sigma^\dagger)_3^{~1}
		(\Sigma)_2^{~1}(j/j_5)$
		\\
		&$\tilde{\calO}_{udus}^{LRRL,S/P}$
		& $ {\bf 8}_L\times {\bf 8}_R(b) $
		& $g_{8\times 8}^b\frac{F_0^4}{4}(\Sigma^\dagger)_2^{~1}
		(\Sigma)_3^{~1}(j/j_5)$
		\\
		&$\tilde{\calO}_{usud}^{LRRL,S/P}$
		& $ {\bf 8}_L\times {\bf 8}_R(b) $
		& $g_{8\times 8}^b\frac{F_0^4}{4}(\Sigma^\dagger)_3^{~1}
		(\Sigma)_2^{~1}(j/j_5)$
		\\
		\hline
	\end{tabular}
	\caption{Chiral realizations (fourth column) of dim-9 LEFT operators (second column) contributing to decays $\tau^+\to\ell^-P^+_iP^+_j$ with $\ell=e,~\mu$; $P_i=\pi,~K$ for $i=2,3$; and $j_{(5)}=j^{\ell\tau}_{(5)}$ or $j^{\tau\ell}_{(5)}$.}
	\label{tabLEFT2chpt}
\end{table}

Now we turn to evaluate the SD contribution in Fig.~\ref{fig1}~(c) that arises from the matching with the dim-9 LEFT operators in Table~\ref{tabSM2LEFT}. We refer to our previous work~\cite{Liao:2019gex} for details of matching based on spurion analysis, and we show the results in Table~\ref{tabLEFT2chpt}. This matching leaves behind a low energy constant (LEC), which is multiplied by a mesonic operator that can only be determined by nonperturbative methods. Note that different components in the same irreducible representation of the chiral group share the same LEC in the chiral limit; for instance, all of $\calO_{udud}^{LLLL,S/P}$, $\calO_{usus}^{LLLL,S/P}$, and $\calO_{udus}^{LLLL,S/P}$ share the same LEC $g_{27\times 1}$. Operators belonging to the same irreducible representation but arising from different color contractions generally have different LECs, as is the case with the LECs $g_{8\times 8}^a$ and $g_{8\times 8}^b$. Fortunately, these three parameters are already determined in the literature; here, we use the values from~\cite{Cirigliano:2017djv}, which in our notation are
\begin{align}
	&g_{27\times 1}=0.38\pm 0.08,~
	&&g_{8\times 8}^a=5.5\pm2~{\GeV}^2,~
	&&g_{8\times 8}^b=1.55\pm0.65~{\GeV}^2.
	\label{lecs}
\end{align}
Expanding the hadronic operators in Table~\ref{tabLEFT2chpt} to their first terms and attaching their corresponding LEFT Wilson coefficients defined at the scale $\Lambda_\chi$ yields the SD interactions for $\ell_\alpha^\pm\ell_\beta^\pm P_i^\mp P_j^\mp$:
\begin{align}
	\label{lshortchi}
	\calL^\textrm{SD}_{\ell_\alpha^\pm\ell_\beta^\pm P_i^\mp P_j^\mp}=&
\frac{5}{6}F_0^2 g_{27\times 1}\partial^\mu P_i^{-}\partial_\mu P_j^{-}\left[
	C^{LLLL,S}_{uiuj}(\Lambda_\chi)\overline{\ell_{\alpha}}\ell^C_{\beta}
	+C^{LLLL,P}_{uiuj}(\Lambda_\chi)
	\overline{\ell_{\alpha}}\gamma_5\ell^C_{\beta}\right]
	\nonumber
	\\
	&+\frac{1}{2}F_0^2P_i^-P_j^- \left[\left(C^{LRRL,S}_{uiuj}
	(\Lambda_\chi)g_{8\times8}^a
	+\tilde{C}^{LRRL,S}_{uiuj}(\Lambda_\chi)g_{8\times8}^{b}\right)
	\overline{\ell_{\alpha}}\ell^C_{\beta}\right.
	\nonumber
	\\
	&+\left.\left(C^{LRRL,P}_{uiuj}(\Lambda_\chi)g_{8\times8}^{a}
	+\tilde{C}^{LRRL,P}_{uiuj}(\Lambda_\chi)g_{8\times8}^{b}\right)
	\overline{\ell_{\alpha}}\gamma_5\ell^C_{\beta} +(1-\delta_{ij})(i\leftrightarrow j)\right]
+\textrm{h.c.}.
\end{align}
Utilizing the QCD running effects [Eqs.~\eqref{eqr1}-\eqref{eqr3}] and the LEFT--SMEFT matching results in Table~\ref{tabSM2LEFT}, we obtain
\begin{align}
	\mathcal{L}_{\tau^+\to\ell^-P_i^+P_j^+}^{\rm SD}= {2F_0^2G_F\over 1+\delta_{ij} }\left[
	c_{1,ij}^{\ell\tau}P_i^{-}P_j^{-}
	+c_{5,ij}^{\ell\tau}\partial^\mu P_i^{-}\partial_\mu P_j^{-}\right]\overline{\ell_{L}}\tau^C_{L},
\end{align}
where the parameters $c_{1,5}$ are defined as
\begin{align}
	\label{wc1}
	&c_{1,ij}^{\ell\tau}=
	-2\sqrt{2}\left(0.62 g_{8\times8}^{a}+0.88 g_{8\times8}^{b}\right)
	\calX^{\ell\tau}_{1,P_iP_j},%\\
	&&c_{5,ij}^{\ell\tau}=
	-2\sqrt{2}(1.3g_{27\times 1})V_{ui}V_{uj}
	\calX^{\ell\tau}_{2}.
\end{align}
Here,
\begin{align}	
	\calX^{\ell\tau}_{1,P_iP_j}=&
	2V_{ui}C_{\bar{d}uLDL}^{ju\ell\tau*}(\Lambda_{\rm EW})
	+2V_{uj}C_{\bar{d}uLDL}^{iu\ell\tau*}(\Lambda_{\rm EW}), \nonumber
	\\
	\calX^{\ell\tau}_{2}=&
	2C_{LHW}^{\ell\tau*}(\Lambda_{\rm EW})
	+2C_{LHW}^{\tau\ell*}(\Lambda_{\rm EW})
	+2C_{LDH1}^{\ell\tau*}(\Lambda_{\rm EW})
	+C_{LDH2}^{\ell\tau*}(\Lambda_{\rm EW}),
\label{Xdefined}
\end{align}
are given in terms of the SMEFT Wilson coefficients evaluated at the electroweak scale $\Lambda_{\rm EW}$.

With all relevant interactions between the mesons and leptons at hand, we can compute the Feynman diagrams in Fig.~\ref{fig1} to obtain the complete amplitude for the decay $\tau^+(p_1)\rightarrow \ell^-(p_2)P_i^{+}(q_1)P_j^{+}(q_2)$,
\begin{align}
	\label{amplitude1}
	\mathcal{M}=F_0^2G_F\left[ T_\textrm{SD}\overline{v_\tau}P_Ru_\ell^C
	+T_{1\mu\nu}\overline{v_\tau}\gamma^\mu\gamma^\nu P_Ru_\ell^C
	+T_{2\mu\nu\rho}\overline{v_\tau}\gamma^\mu\gamma^\nu\gamma^\rho P_Ru_\ell^C
	+T_{3\mu\nu\rho}\overline{v_\tau}\gamma^\mu\gamma^\nu\gamma^\rho P_Lu_\ell^C
	\right],
\end{align}
where $T_\textrm{SD}$ denotes the SD term and the others are the LD ones; that is,
\begin{align}
	T_\textrm{SD}=&-2\left(c_{1,ij}^{\tau\ell}-c_{5,ij}^{\tau\ell} (q_1\cdot q_2)\right),
	\\
	\nonumber
	T_{1\mu\nu}=&G_FV_{ui}V_{uj}m_{\tau\ell}\left( q_{1\mu} q_{2\nu} t^{-1}
	+q_{2\mu} q_{1\nu} u^{-1}
	\right)
	\\
	\nonumber
	&+\left[V_{ui}\left( Bc_{P_j1}^{\ell\tau}-c_{P_j3}^{\ell\tau}(t-p_2^2)  \right)q_{1\mu}(p_1-q_1)_\nu-V_{uj}\left(Bc_{P_i1}^{\tau\ell}-c_{P_i3}^{\tau\ell}(t-p_1^2) \right)(p_1-q_1)_\mu q_{2\nu}\right]t^{-1}
	\\
	&+\left[V_{uj}\left(Bc_{P_i1}^{\ell\tau}-c_{P_i3}^{\ell\tau}(u-p_2^2)\right) q_{2\mu}(p_1-q_2)_\nu-V_{ui}\left(Bc_{P_j1}^{\tau\ell}-c_{P_j3}^{\tau\ell} (u-p_1^2)\right)(p_1-q_2)_\mu q_{1\nu} \right]u^{-1},
	\\
	T_{2\mu\nu\rho}=&V_{uj}c_{P_i2}^{\tau\ell}q_{1\mu}(p_1-q_1)_\nu q_{2\rho} t^{-1}
	+V_{ui}c_{P_j2}^{\tau\ell}q_{2\mu} (p_1-q_2)_\nu q_{1\rho} u^{-1},
	\\
	T_{3\mu\nu\rho}=&V_{ui}c_{P_j2}^{\ell\tau}q_{1\mu}(p_1-q_1)_\nu q_{2\rho} t^{-1}
	+ V_{uj}c_{P_i2}^{\ell\tau}q_{2\mu}(p_1-q_2)_\nu q_{1\rho} u^{-1},
\end{align}
with $s=(q_1+q_2)^2,~t=(p_1-q_1)^2$, and $u=(p_1-q_2)^2$.

\section{Discussions of hadronic uncertainties and improvement}
\label{disp}

So far, we have been working to the leading order in $\chpt$. It is well-known that chiral perturbation does not converge fast enough for hadronic $\tau$ decays owing to its large mass $m_\tau$ compared to $\Lambda_\chi$; thus, in this section, we estimate the uncertainties due to ignored higher order corrections and seek to improve our leading order results. To estimate uncertainties, we compute chiral logarithms arising from the one-loop diagrams associated with an LNV vertex, which (as in the usual case) cannot be cancelled by higher order counterterms. Futhermore, to improve convergence we employ the dispersion relation technique, by incorporating experimental data regarding phase shifts. Inspection of Fig.~\ref{fig1} shows that this is nontrivial only for the SD part, which unfortunately is numerically less significant than the LD one, as will be seen in Section~\ref{constr}. In short, we are effectively considering one-loop uncertainties or dispersion-relation improvements to the matrix element $\langle P_i^+(q_1) P_j^+(q_2)|\calO_\textrm{irrep}(0)|0\rangle$, where $\calO_\textrm{irrep}$ is a dim-9 LEFT operator in the chiral representation irrep in Table~\ref{tabLEFT2chpt}, whose lepton bilinear has been stripped off for simplicity and whose chiral realization is also shown in the table.

\begin{figure}%[!h]
	\centering.
	\includegraphics[width=10cm]{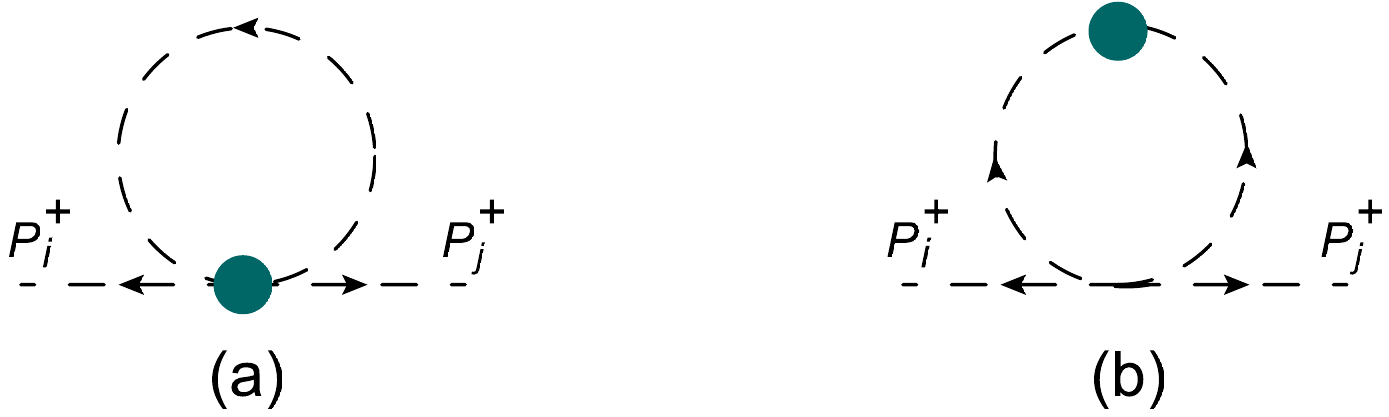}
	\caption{One-loop diagrams for SD contributions to $\langle P_i^+P_j^+|\calO_\textrm{irrep}|0\rangle$ with an insertion of $\calO_\textrm{irrep}$ in shaded circle}.
	\label{figSDlog}
\end{figure}

To assess the relative importance of one-loop chiral logarithms to tree-level terms, we compute the one-loop diagrams in Fig.~\ref{figSDlog} at the kinematic point $(q_1+q_2)^2=0$, as seen in~\cite{Liao:2019gex} for LNV $K^\pm$ decays. The results are
\begin{eqnarray}
	\mathcal{M}_{27\times 1}^{\pi\pi}&=&
	\frac{5}{3}g_{27\times 1}F_\pi^2 m_\pi^2\left(1+3L_\pi\right),
	\\
	\mathcal{M}_{8\times 8}^{\pi\pi,a/b}&=&
	g_{8\times 8}^{a/b}F_\pi^2\left(1+L_\pi\right),
	\\
	\mathcal{M}_{27\times 1}^{KK}&=&
	\frac{5}{3}g_{27\times 1}F_K^2 m_K^2\left[1+\frac{1}{4}\left(\frac{m_K^2+ m_\pi^2}{ m_K^2}L_\pi+4L_K+\frac{7m_K^2-m_\pi^2}{m_K^2}L_\eta\right)\right],
	\\
	\mathcal{M}_{8\times 8}^{KK,a/b}&=&
	g_{8\times 8}^{a/b}F_K^2\left[1-\frac{1}{4}\left(L_\pi-8L_K+3L_\eta\right)\right],
	\\
	\mathcal{M}_{27\times 1}^{K\pi}&=&
	\frac{5}{12}g_{27\times 1}F_K^2(m_K^2+m_\pi^2)\left[1-\frac{1}{4}
	\left(\frac{17m_\pi^2-9m_K^2}{2(m_K^2-m_\pi^2)}L_\pi
	-\frac{5m_K^2-m_\pi^2}{m_K^2-m_\pi^2}L_K+\frac{3}{2}L_\eta\right)\right],
	\\
	\mathcal{M}_{8\times 8}^{K\pi,a/b}&=&
	\frac{1}{2}g_{8\times 8}^{a/b}F_K^2\left[1
	-\frac{1}{4}\left(\frac{9m_\pi^2-m_K^2}{2(m_K^2-m_\pi^2)}L_\pi
	-\frac{m_K^2+3m_\pi^2}{m_K^2-m_\pi^2}L_K
	+\frac{3}{2}L_\eta\right)\right],
\end{eqnarray}
where $L_P=m_P^2/(4\pi F_0)^2\ln(\mu^2/m_P^2)$ with $\mu$ being the renormalization scale. The results for the $K\pi$ channel coincide with those in~\cite{Liao:2019gex}, whilst those for the $KK$ and $\pi\pi$ channels are newly computed. We have taken into account both renormalization of decay constants and wavefunction renormalization collected in~\cite{Liao:2019gex}. As a rough estimate, the relative corrections at $\mu=\Lambda_\chi$ ($\mu=m_\tau$) in the $\pi\pi$, $K\pi$, and $KK$ channels (each placed in a pair of square brackets) and in the order of the chiral representations
${\bf 27}_L\times{\bf 1}_R$, ${\bf 8}_L\times{\bf 8}_R(a/b)$ (separated by a comma within a pair of square brackets) in each channel are,
[$27\%$, $17\%$] ([$29\%$, $18\%$]),
[$50\%$, $28\%$] ([$55\%$, $27\%$]),
[$65\%$, $50\%$] ([$73\%$, $56\%$]), respectively. The neglected higher order corrections are thus about $20\%-70\%$.

Here, we improve the leading order terms using the dispersion relation technique. For its recent application to $\tau$ decays, see,~\cite{Daub:2012mu,Rendon:2019awg,
Cirigliano:2017tqn,Miranda:2018cpf,Garces:2017jpz,Celis:2013xja}. The aforementioned  matrix elements are parameterized as
\begin{eqnarray}
\calM_{27\times 1}^{P_i P_j}(s)&=&
\langle P_i^+(q_1) P_j^+(q_2)|(\overline{u_L}\gamma^\mu d_L^i)
[\overline{u_L}\gamma_\mu d_L^j]|0\rangle
=-(q_1\cdot q_2)F^{P_i P_j}_{27\times 1}(s)(1+\delta_{ij}),
\\
\calM_{8\times 8}^{P_i P_j,a}(s)&=&
\langle P_i^+(q_1) P_j^+(q_2)|(\overline{u_L} d_R^i)
[\overline{u_R} d_L^j]|0\rangle
=F^{P_i P_j,a}_{8\times 8}(s)(1+\delta_{ij}),
\\
\calM_{8\times 8}^{P_i P_j,b}(s)&=&
\langle P_i^+(q_1) P_j^+(q_2)|(\overline{u_L} d_R^i]
[\overline{u_R} d_L^j)|0\rangle
=F^{P_i P_j,b}_{8\times 8}(s)(1+\delta_{ij}),
\end{eqnarray}
with $s=(q_1+q_2)^2$. The form factors are, to leading order in $\chpt$, normalized to
\begin{align}
&F_{27\times 1}^{P_iP_j}(0)=\frac{5}{6}F_0^2 g_{27\times 1},
&&F_{8\times 8}^{P_iP_j,a/b}(0)
=\frac{1}{2}F_0^2g_{8\times 8}^{~a/b}.
\end{align}

\begin{figure}%[!h]
\centering
\includegraphics[width=16cm]{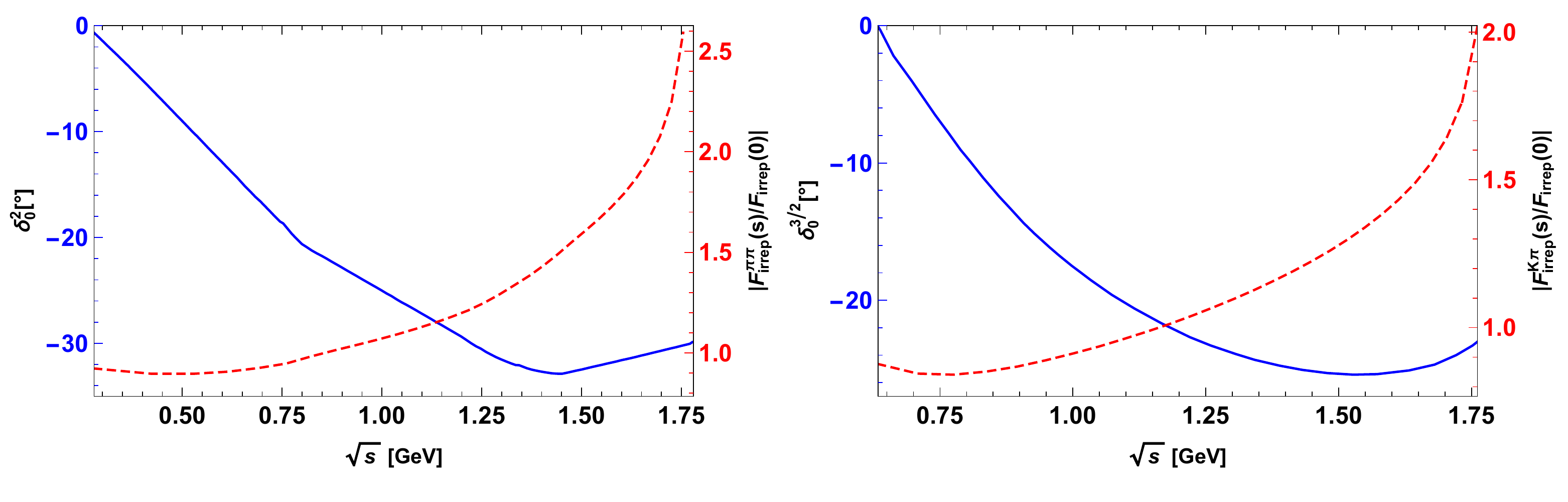}%
\caption{Phase shifts (blue/solid curve, left vertical axis) and $|F^{P_iP_j}_\textrm{irrep}(s)/F^{P_iP_j}_\textrm{irrep}(0)|$ (red/dashed curve, right vertical axis) in the $\pi\pi$ (left panel) and $K\pi$ (right panel) channels, displayed as a function of $\sqrt{s}$.}
\label{Disppipi}
\end{figure}

To construct the dispersion relation, we decompose the elastic meson scattering amplitude into partial wave amplitudes $f_l^I(s)$ with orbital angular momentum $l$ and isospin $I$. Application of the Cutkosky cutting rules to Fig.~\ref{figSDlog}~(b) (where the meson scattering vertex is replaced by a general scattering amplitude), yields
\begin{eqnarray}
	\label{Fpipidisp}
	{\rm Im}~F_\textrm{irrep}^{\pi\pi}(s)&=&
\frac{2 \lambda^{1/2}_{\pi\pi}(s)}{s}
F_\textrm{irrep}^{\pi\pi}(s)[f_0^2(s)]^*\theta(s-s_{\pi\pi}),
\\
	\label{Fkkdisp}
	{\rm Im}~F_\textrm{irrep}^{KK}(s)&=&
\frac{2 \lambda^{1/2}_{KK}(s)}{s}
F_\textrm{irrep}^{KK}(s)[f_0^1(s)]^*\theta(s-s_{KK}),
\\
	\label{Fkpidisp}
	{\rm Im}~F_\textrm{irrep}^{K\pi}(s)&=&
\frac{\lambda^{1/2}_{K\pi}(s)}{s}
F_\textrm{irrep}^{K\pi}(s)[f_0^{3/2}(s)]^*\theta(s-s_{K\pi}),
\end{eqnarray}
where $s_{P_i P_j}=(m_{P_i}+m_{P_j})^2$, and $\lambda_{P_i P_j}(s)$ is the basic three-particle kinematic function:
\begin{eqnarray}
\lambda_{P_i P_j}(s)=m_{P_i}^2+m_{P_j}^2+s-2m_{P_i}^2 m_{P_j}^2-2m_{P_i}\sqrt s-2m_{P_j}\sqrt s.
\end{eqnarray}
The partial wave amplitude for elastic scattering can be expressed in terms of the phase shift $\delta_l^I(s)$:
\begin{equation}\label{partialwave}
	f_l^I(s)=\frac{s}{\lambda^{1/2}(s)}\sin \delta_l^I(s)e^{i\delta_l^I(s)}.
\end{equation}
The above dispersion relations imply that the phase of $F_\textrm{irrep}^{P_{i}P_{j}}(s)$ is equal to the corresponding phase shift, similar to the Watson's final-state theorem for elastic scattering states~\cite{Watson:1954uc}. Eqs.~\eqref{Fpipidisp}-\eqref{Fkpidisp} then have a universal solution of
\begin{align}
F_\textrm{irrep}^{P_i P_j}(s)
=F_\textrm{irrep}^{P_i P_j}(0)\Omega^I_l(s),
\end{align}
where the Omn\`es factor $\Omega^I_l(s)$~\cite{Omnes:1958hv} for a once-subtracted dispersion relation is
\begin{align}
\Omega^I_l(s)=\exp\left[
\frac{s}{\pi}\int_{s_{P_i P_j}}^\infty ds'
\frac{\delta^I_l(s')}{s'(s'-s)}\right].
\end{align}
The phase shifts $\delta_0^2$ and $\delta_0^{3/2}$ have been measured experimentally~\cite{Losty:1973et,Hoogland:1977kt}. We have taken the fits of $\delta_0^2$ from~\cite{Descotes-Genon:2014tla} and of $\delta_0^{3/2}$ from~\cite{Rodas:2019ofu}. 
To the best of our knowledge, no data are available for the $(I,l)=(1,0)$ channel. These phase shifts and corresponding magnitudes of $F_\textrm{irrep}^{P_i P_j}(s)/F_\textrm{irrep}^{P_i P_j}(0)=\Omega^I_l(s)$ are shown in Fig.~\ref{Disppipi}, in which a cutoff $s_{\rm cut}=m_{\tau}^2$ has been chosen for the integral. As expected, no resonance is indicated in these channels. The above results are applied to our decay rate evaluation in the next section.

%%%%%%%%%%%%%%%
\section{Master formulas for decay rates}
\label{constr}
%%%%%%%%%%%%%%%

Here, we present our master formulas for the decay rates and branching ratios of the $\tau^+\to \ell^-P_i^+P_j^+$ decay. We omit the kinematic details because they are similar to the LNV $K^+$ decays~\cite{Liao:2019gex,Liao:2020roy}. The spin-summed and -averaged decay width is
\begin{align}
\Gamma=\frac{1}{1+\delta_{ij}}\frac{1}{2m_\tau}
\frac{1}{128\pi^3m_\tau^2}
\int ds \int dt~ \overline{\sum}|\mathcal{M}|^2,
\label{decaywidth}
\end{align}
for which the integration domains are
\begin{align}
&s\in \left[(m_{P_i}+m_{P_j})^2,~(m_\tau-m_\ell)^2\right],
\\
&t\in \Big[(E_2^*+E_3^*)^2-\Big(\sqrt{E_2^{*2}-m_{P_j}^2}+\sqrt{E_3^{*2}-m_\ell^2}
\Big)^2,
\nonumber
\\
&\hspace{0.65cm}(E_2^*+E_3^*)^2-\Big(\sqrt{E_2^{*2}-m_{P_j}^2}
-\sqrt{E_3^{*2}-m_\ell^2}\Big)^2\Big],
\end{align}
with
\begin{align}
&E_2^*=\frac{1}{2\sqrt{s}}(s-m_{P_i}^2+m_{P_j}^2),
&&E_3^*=\frac{1}{2\sqrt{s}}(m_\tau^2-s-m_\ell^2).
\end{align}
Using the LECs in Eq.~\eqref{lecs} and the SM parameters for the $\tau$ lepton width, various particle masses, and the Fermi constant $G_F$~\cite{Zyla:2020zbs}, the decay branching ratios become
\begin{align}
\nonumber
{\mathcal{B}(\tau^+\rightarrow e^-\pi^+\pi^+)\over  \GeV^6}=&
\frac{2.4\times 10^{-34}}{\GeV^6}\frac{|m_{\tau e}|^2}{\eV^2}
+0.31\left|\calY^{e\tau}_{\pi 1}\right |^2
+0.21\left|\calY^{\tau e}_{\pi 1}\right |^2
+1.9\times10^{-3}\left(\left|\calY^{\tau e}_{\pi 2}\right|^2+\left|\calY^{e\tau}_{\pi 2}\right|^2\right)
\\
\label{rpipie}
&
+5.8\times10^{-4}\left|\calX^{\tau e}_{1,\pi\pi}\right|^2
+10^{-8}\left(58\left|\calX^{\tau e}_{2}\right|^2
+33\left|\calY^{\tau e}_{\pi 3}\right|^2
+2.2\left|\calY^{e\tau}_{\pi 3}\right|^2\right)+\textrm{int.},
\\
\nonumber
{\mathcal{B}(\tau^+\rightarrow \mu^-\pi^+\pi^+)\over  \GeV^6}=&
\frac{8.1\times 10^{-35}}{\GeV^6}\frac{|m_{\tau\mu}|^2}{\eV^2}
+0.26\left|\calY^{\mu\tau}_{\pi 1}\right |^2
+0.19\left|\calY^{\tau \mu}_{\pi 1}\right |^2
+1.7\times10^{-3}\left(\left|\calY^{\tau \mu}_{\pi 2}\right|^2
+\left|\calY^{\mu\tau}_{\pi 2}\right|^2\right)
\\
\label{rpipimu}
&
+5.6\times10^{-4}\left|\calX^{\tau \mu}_{1,\pi\pi}\right|^2
+10^{-8}\left(53\left|\calX^{\tau \mu}_{2}\right|^2
+29\left|\calY^{\tau \mu}_{\pi 3}\right|^2
+2.2\left|\calY^{\mu\tau}_{\pi 3}\right|^2\right)+\textrm{int.},
\\
\nonumber
{\mathcal{B}(\tau^+\rightarrow e^-K^+K^+)\over  \GeV^6}=&
\frac{3.1\times 10^{-38}}{\GeV^6}\frac{|m_{\tau e}|^2}{\eV^2}
+2.3\times10^{-3}\left|\calY^{\tau e}_{K 1}\right |^2
+1.5\times10^{-3}\left|\calY^{e\tau}_{K 1}\right |^2
+1.0\times10^{-4}\left|\calX^{\tau e}_{1,KK}\right|^2
\\
\nonumber
&
+2.1\times10^{-5}\left(\left|\calY^{\tau e}_{K 2}\right|^2+\left|\calY^{e\tau}_{K 2}\right|^2\right)
\\
\label{rkke}
&+10^{-9}\left(4.2\left|\calY^{\tau e}_{K 3}\right|^2
+0.34\left|\calY^{e\tau}_{K 3}\right|^2
+0.23\left|\calX^{\tau e}_{2}\right|^2\right)+\textrm{int.},
\\
\nonumber
{\mathcal{B}(\tau^+\rightarrow \mu^-K^+K^+)\over  \GeV^6}=&
\frac{2.5\times 10^{-38}}{\GeV^6}\frac{|m_{\tau\mu}|^2}{\eV^2}
+2.1\times10^{-3}\left|\calY^{\tau \mu}_{K 1}\right |^2
+1.3\times10^{-3}\left|\calY^{\mu\tau}_{K 1}\right |^2
+9.9\times10^{-5}\left|\calX^{\tau \mu}_{1,KK}\right|^2
\\
\nonumber
&
+1.9\times10^{-5}\left(\left|\calY^{\tau \mu}_{K 2}\right|^2
+\left|\calY^{\mu\tau}_{K 2}\right|^2\right)
\\
\label{rkkmu}
&+10^{-9}\left(3.7\left|\calY^{\tau \mu}_{K 3}\right|^2
+0.33\left|\calY^{\mu\tau}_{K 3}\right|^2
+0.25\left|\calX^{\tau \mu}_{2}\right|^2\right)+\textrm{int.},
\\
\nonumber
{\mathcal{B}(\tau^+\rightarrow e^-K^+\pi^+)\over  \GeV^6}=&
\frac{6.2\times 10^{-36}}{\GeV^6}\frac{|m_{\tau e}|^2}{\eV^2}
+5.5\times10^{-2}\left|\calY^{\tau e}_{K 1}\right |^2
+4.8\times10^{-2}\left|\calY^{e\tau}_{K 1}\right |^2
+1.5\times10^{-2}\left|\calY^{e\tau}_{\pi 1}\right |^2
\\
\nonumber
&
+10^{-5}\left(290\left|\calY^{\tau e}_{\pi 1}\right|^2
+93\left|\calY^{\tau e}_{K 2}\right|^2
+48\left|\calY^{e\tau}_{K 2}\right|^2
+13\left|\calX^{\tau e}_{1,K\pi}\right|^2
+5.0\left|\calY^{e\tau}_{\pi 2}\right|^2
+2.6\left|\calY^{\tau e}_{\pi 2}\right|^2\right)
\\
\label{rkpie}
&
+10^{-9}\left(130\left|\calY^{\tau e}_{K 3}\right|^2
+26\left|\calX^{\tau e}_{2}\right|^2
+17\left|\calY^{e\tau}_{K 3}\right|^2
+3.6\left|\calY^{\tau e}_{\pi 3}\right|^2
+0.9\left|\calY^{e\tau}_{\pi 3}\right|^2\right)+\textrm{int.},
\\
\nonumber
{\mathcal{B}(\tau^+\rightarrow \mu^-K^+\pi^+)\over  \GeV^6}=&
\frac{4.2\times 10^{-36}}{\GeV^6}\frac{|m_{\tau \mu}|^2}{\eV^2}
+5\times10^{-2}\left|\calY^{\tau \mu}_{K 1}\right |^2
+4.3\times10^{-2}\left|\calY^{\mu\tau}_{K 1}\right |^2
+1.2\times10^{-2}\left|\calY^{\mu\tau}_{\pi 1}\right |^2
\\
\nonumber
&
+10^{-5}\Big(280\left|\calY^{\tau \mu}_{\pi 1}\right|^2
+80\left|\calY^{\tau \mu}_{K 2}\right|^2
+43\left|\calY^{\mu\tau}_{K 2}\right|^2
+12\left|\calX^{\tau \mu}_{1,K\pi}\right|^2
+4.3\left|\calY^{\mu\tau}_{\pi 2}\right|^2
\\
\nonumber
&
+2.3\left|\calY^{\tau \mu}_{\pi 2}\right|^2\Big)
+10^{-9}\left(110\left|\calY^{\tau \mu}_{K 3}\right|^2
+24\left|\calX^{\tau \mu}_2\right|^2
+16\left|\calY^{\mu\tau}_{K 3}\right|^2
+3.3\left|\calY^{\tau \mu}_{\pi 3}\right|^2\right)
\\
\label{rkpimu}
&+8.5\times10^{-10}\left|\calY^{\mu\tau}_{\pi 3}\right|^2+\textrm{int.},
\end{align}
where the Wilson coefficients of the dim-7 SMEFT operators are contained in the $\calX$ (for the SD part) and $\calY$ (for the LD part) parameters defined in Eqs.~\eqref{Xdefined} and \eqref{Ydefined}, for which the interference terms (int.) are not explicitly displayed. We have incorporated the dispersion-relation-improved hadronic matrix elements into the SD part.

\begin{figure}
	\centering
	\includegraphics[width=10cm]{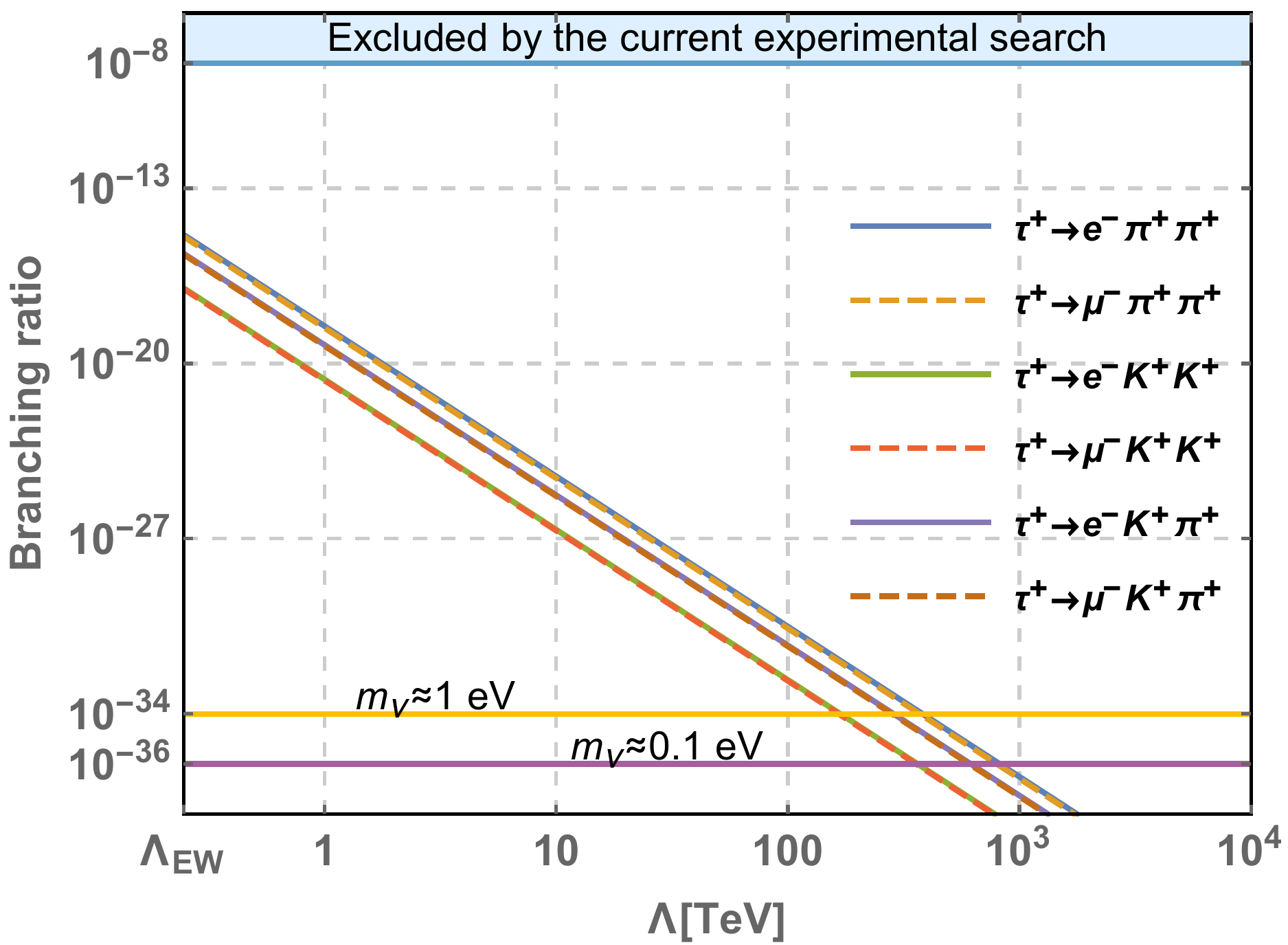}
	\caption{Branching ratios for $\tau^+\rightarrow \ell^-P_i^{+}P_j^{+}$ shown as a function of the new physics scale $\Lambda$ under the assumption of identical Wilson coefficients $C_i=\Lambda^{-3}$ for dim-7 operators in SMEFT. The upper horizontal line denotes current experimental bounds and the lower ones denote the neutrino mass contribution alone.}
	\label{FigPredict}
\end{figure}

The above results show that the contribution from neutrino mass insertion in Fig.~\ref{fig1}~(a) can be entirely neglected. If we assume that the Wilson coefficients of LNV dim-7 operators in the SMEFT are of a similar size, their relative importance is then controlled by the prefactors of the $\calX$ and $\calY$ parameters. The LD contribution from $\calY_{1}$ dominates, whilst those from $\calY_{2}$ and $\calY_{3}$ are suppressed by factors of $p$ and $p/\Lambda_{\rm EW}$, respectively. The SD contribution of $\calX_{1}$ has an order of magnitude similar to $\calY_{2}$, whilst the $\calX_{2}$ term is suppressed by $p^2/\Lambda_\chi^2$ and has a similar size to $\calY_{3}$. To obtain concrete constraints, we must make a simplifying assumption, because there are too many Wilson coefficients; hence, we assume that only one of the $\calX$ and $\calY$ parameters is nonzero at a time. The experimental upper bounds on the $\tau$ decays in Eqs.~\eqref{ul1}-\eqref{ul3} translate to the bounds on those parameters, as shown in Table~\ref{tablbNP}. These bounds are significantly weaker than those from the nuclear $0\nu\beta\beta$ decay and LNV $K^\pm$ decays, owing to the much smaller data samples; and being of order GeV they should not be taken literally. But they are the first bounds obtained thus far for the LNV Wilson coefficients in the SMEFT that involve the third generation of leptons, and they are comparable to those that would be expected to set at the LHC on the $\mu\mu$ component of the Weinberg operator, (see~\cite{Fuks:2020zbm} for a recent discussion). If we parameterize all Wilson coefficients by the same scale $C_i=\Lambda^{-3}$, the branching ratios will be proportional to $\Lambda^{-6}$, as shown in Fig.~\ref{FigPredict}. For $\Lambda>1~\TeV$, we have
\begin{eqnarray}
\mathcal{B}(\tau^-\rightarrow e^+\pi^-\pi^-)<3.1\times10^{-19},&\mathcal{B}(\tau^-\rightarrow \mu^+ \pi^-\pi^-)<2.6\times10^{-19},\\
\mathcal{B}(\tau^-\rightarrow e^+K^-K^-)<2.3\times10^{-21},&\mathcal{B}(\tau^-\rightarrow \mu^+ K^-K^-)<2.1\times10^{-21},\\
\mathcal{B}(\tau^-\rightarrow e^+K^-\pi^-)<5.5\times10^{-20},&\mathcal{B}(\tau^-\rightarrow \mu^+ K^-\pi^-)<5.0\times10^{-20},
\end{eqnarray}
which are several orders of magnitude smaller than the current experimental upper bounds.

\begin{table}[t]
\centering
\begin{tabular}{| lc| lc| lc| lc|}
\hline
\multicolumn{2}{|c}{$\tau^+\rightarrow e^-\pi^+\pi^+ $} &  \multicolumn{2}{|c}{ $\tau^+\rightarrow e^-K^+K^+ $}
&\multicolumn{4}{|c|}{ $\tau^+\rightarrow e^-K^+\pi^+ $}
\\
\hline
Name & Bounds &Name & Bounds &Name & Bounds &Name & Bounds
\\
\hline
$\left|\calY^{e\tau}_{\pi1}\right |^{-\frac{1}{3}}$ & $15.8$
& $\left|\calY^{\tau e}_{K1}\right |^{-\frac{1}{3}}$ & $6.4$
& $\left|\calY^{\tau e}_{K1}\right |^{-\frac{1}{3}}$ & $10.9$
& $\left|\calY^{e \tau}_{K1}\right |^{-\frac{1}{3}}$ & $10.7$
 \\
\hline
$\left|\calY^{\tau e}_{\pi1}\right |^{-\frac{1}{3}}$ & $14.8$
& $\left|\calY^{e\tau}_{K1}\right |^{-\frac{1}{3}}$ &  $6.0$
& $\left|\calY^{e\tau}_{\pi1}\right |^{-\frac{1}{3}}$ & $8.8$
& $\left|\calY^{\tau e}_{\pi1}\right |^{-\frac{1}{3}}$ & $6.7$
 \\
 \hline
$\left|\calY^{e\tau}_{\pi2}\right |^{-\frac{1}{3}}$ & $6.8$
& $\left|\calX^{\tau e}_{1,KK}\right |^{-\frac{1}{3}}$ & $3.8$
& $\left|\calY^{\tau e}_{K2}\right |^{-\frac{1}{3}}$ & $5.5$
& $\left|\calY^{e\tau}_{K2}\right |^{-\frac{1}{3}}$ & $5.0$
\\
\hline
$\left|\calY^{\tau e}_{\pi2}\right |^{-\frac{1}{3}}$ & $6.8$
&$\left|\calY^{\tau e}_{K2}\right |^{-\frac{1}{3}}$ &  $2.9$
&$\left|\calX^{\tau e}_{1,K\pi}\right |^{-\frac{1}{3}}$ & $4.0$
& &
 \\
 \hline
$\left|\calX^{\tau e}_{1,\pi\pi}\right |^{-\frac{1}{3}}$ & $5.5$
& $\left|\calY^{e\tau}_{K2}\right |^{-\frac{1}{3}}$ & $2.9$
& $\left|\calY^{e\tau}_{\pi2}\right |^{-\frac{1}{3}}$& $3.4$
& $\left|\calY^{\tau e}_{\pi2}\right |^{-\frac{1}{3}}$& $3.1$
 \\
 \hline
$\left|\calX^{\tau e}_2\right |^{-\frac{1}{3}}$ & $1.8$
& $\left|\calY^{\tau e}_{K3}\right |^{-\frac{1}{3}}$ & $0.7$
& $\left|\calY^{\tau e}_{K3}\right |^{-\frac{1}{3}}$& $1.3$
& $\left|\calY^{e\tau}_{K3}\right |^{-\frac{1}{3}}$& $0.9$
 \\
 \hline
 $\left|\calY^{\tau e}_{\pi 3}\right |^{-\frac{1}{3}}$ & $1.6$
 &  $\left|\calY^{e\tau}_{K3}\right |^{-\frac{1}{3}}$ & $0.5$
 & $\left|\calX^{\tau e}_{2}\right |^{-\frac{1}{3}}$ & $1.0$
&  &
 \\
 \hline
  $\left|\calY^{e\tau}_{\pi3}\right |^{-\frac{1}{3}}$ & $1.0$
  & $\left|\calX^{\tau e}_2\right |^{-\frac{1}{3}}$ &  $0.4$
  & $\left|\calY^{\tau e}_{\pi 3}\right |^{-\frac{1}{3}}$ & $0.7$
  & $\left|\calY^{e\tau }_{\pi 3}\right |^{-\frac{1}{3}}$& $0.6$
  \\
  \hline
  \hline
\multicolumn{2}{|c}{$\tau^+\rightarrow \mu^-\pi^+\pi^+ $} &  \multicolumn{2}{|c}{ $\tau^+\rightarrow \mu^-K^+K^+ $}
&\multicolumn{4}{|c|}{ $\tau^+\rightarrow \mu^-K^+\pi^+ $}
\\
\hline
Name & Bounds &Name & Bounds &Name & Bounds &Name & Bounds
\\
\hline
$\left|\calY^{\mu\tau}_{\pi1}\right |^{-\frac{1}{3}}$ & $13.7$
& $\left|\calY^{\tau \mu}_{K1}\right |^{-\frac{1}{3}}$ & $6.0$
& $\left|\calY^{\tau\mu}_{K1}\right |^{-\frac{1}{3}}$ & $10.1$
& $\left|\calY^{\mu\tau}_{K1}\right |^{-\frac{1}{3}}$ & $9.8$
 \\
\hline
$\left|\calY^{\tau \mu}_{\pi1}\right |^{-\frac{1}{3}}$ & $13.0$
& $\left|\calY^{\mu\tau}_{K1}\right |^{-\frac{1}{3}}$ & $5.5$
& $\left|\calY^{\mu\tau}_{\pi1}\right |^{-\frac{1}{3}}$ & $7.9$
& $\left|\calY^{\tau\mu}_{\pi1}\right |^{-\frac{1}{3}}$ & $6.2$
 \\
 \hline
$\left|\calY^{\mu\tau}_{\pi2}\right |^{-\frac{1}{3}}$ & $5.9$
&  $\left|\calX^{\tau \mu}_{1,KK}\right |^{-\frac{1}{3}}$& $3.6$
&  $\left|\calY^{\tau\mu}_{K2}\right |^{-\frac{1}{3}}$& $5.1$
&  $\left|\calY^{\mu\tau}_{K2}\right |^{-\frac{1}{3}}$& $4.6$
\\
\hline
$\left|\calY^{\tau \mu}_{\pi2}\right |^{-\frac{1}{3}}$ & $5.9$
& $\left|\calY^{\tau\mu}_{K2}\right |^{-\frac{1}{3}}$&  $2.7$
& $\left|\calX^{\tau\mu}_{1,K\pi}\right |^{-\frac{1}{3}}$& $3.7$
& &
 \\
 \hline
$\left|\calX^{\tau \mu}_{1,\pi\pi}\right |^{-\frac{1}{3}}$ & $4.9$
& $\left|\calY^{\mu\tau}_{K2}\right |^{-\frac{1}{3}}$ & $2.7$
& $\left|\calY^{\mu\tau}_{\pi2}\right |^{-\frac{1}{3}}$& $3.1$
& $\left|\calY^{\tau\mu}_{\pi2}\right |^{-\frac{1}{3}}$& $2.8$
 \\
 \hline
$\left|\calX^{\tau \mu}_2\right |^{-\frac{1}{3}}$ & $1.5$
& $\left|\calY^{\tau\mu}_{K3}\right |^{-\frac{1}{3}}$ & $0.7$
& $\left|\calY^{\tau\mu}_{K3}\right |^{-\frac{1}{3}}$& $1.1$
& $\left|\calY^{\mu\tau}_{K3}\right |^{-\frac{1}{3}}$& $0.8$
 \\
 \hline
 $\left|\calY^{\tau \mu}_{\pi 3}\right |^{-\frac{1}{3}}$ & $1.4$
 & $\left|\calX^{\tau\mu}_{2}\right |^{-\frac{1}{3}}$ & $0.4$
 & $\left|\calX^{\tau\mu}_{2}\right |^{-\frac{1}{3}}$ & $0.9$
&  &
 \\
 \hline
  $\left|\calY^{\mu\tau}_{\pi3}\right |^{-\frac{1}{3}}$ & $1.0$
  & $\left|\calY^{\mu\tau}_{K3}\right |^{-\frac{1}{3}}$ & $0.4$
  & $\left|\calY^{\mu\tau}_{\pi3}\right |^{-\frac{1}{3}}$ & $0.6$
&  $\left|\calY^{\tau\mu}_{\pi3}\right |^{-\frac{1}{3}}$& $0.5$
  \\
  \hline
\end{tabular}
\caption{Lower bounds (in units of GeV) on $|\calX_i|^{-1/3}$ or $|\calY_i|^{-1/3}$ parameters for combinations of Wilson coefficients. Note that $\calX_i^{\alpha\beta}=\calX_i^{\beta\alpha}$.}
\label{tablbNP}
\end{table}

%%%%%%%%%%%%%%%
\section{Conclusion}
%%%%%%%%%%%%%%%
\label{conclude}

We studied the LNV $\tau$ decays $\tau^+\rightarrow \ell^-P_i^{+}P_j^{+}$ within the framework of EFT. One merit of these decays is that they could potentially probe LNV interactions in the third generation of leptons, which are not accessible in either nuclear $0\nu\beta\beta$ decay or LNV $K^\pm$ decays. Assuming the absence of new particles of masses below the electroweak scale, we started from the effective interactions of LNV dim-5 and -7 operators in SMEFT; first we matched them to effective interactions in the LEFT at the electroweak scale; then, we matched them to those in $\chpt$ at the chiral symmetry breaking scale. We computed the decay branching ratios and expressed them in terms of the Wilson coefficients in the SMEFT and hadronic low energy constants. As seen in the case of LNV $K^\pm$ decays, the LD contribution from the exchange of a neutrino generically dominates over the SD one arising from LNV dim-9 operators in LEFT involving four quarks and two like-charge leptons. We estimated, by computing one-loop chiral logarithms, the theoretical uncertainties due to neglect of higher order terms in chiral perturbation for the hadronic $\tau$ decays, and found them to be large. Thus, we attempted to improve the convergence in the SD part by appealing to dispersion relations. We found the decays $\tau^+\to e^-\pi^+\pi^+,~\mu^-\pi^+\pi^+$ to have the largest branching ratios among the six channels; however, these are still well below the current experimental bounds for a reasonable choice of a new physics scale.

\vspace{0.5cm}
\noindent %
\section*{Acknowledgement}
%%%%%%%%%%%%%%%%%%%%%%%

This work was supported in part by the Grants No.~NSFC-12035008, No.~NSFC-11975130, by The National Key Research and Development Program of China under Grant No. 2017YFA0402200, by the CAS Center for Excellence in Particle Physics (CCEPP). XDM is supported by the MOST (Grants No. 109-2112-M-002-017-MY3 and 109-2811-M-002-535). We thank Rui Gao and Feng-Kun Guo for electronic communications on the current status of phase shift data.

%%%%%%%%%%%%%%%%%%%%%%%

\end{document}